\documentclass[apl,aps,twocolumn]{revtex4-1}%

\usepackage{amsmath}
\usepackage{graphicx} 
\usepackage{amssymb}
\usepackage{verbatim}
\usepackage[colorlinks=true, pdfstartview=FitV, linkcolor=blue, citecolor=blue, urlcolor=blue]{hyperref} 
\usepackage{pifont}
\usepackage{rotating}

\usepackage{miller}

\def\C        {{$^{13}$C \/}}

\def\NN       {{$^{15}$N \/}}

\newcommand{\Cn}[1]{$^{13}$C$_{#1}$}

\newcommand{\captionstyle}{\normalfont}

\newcommand{\ee}[1]{\cdot10^{#1}}
\newcommand{\mr}[1]{\mathrm{#1}}
\newcommand{\unit}[1]{\,\mathrm{#1}}

\newcommand{\us}{\,\mu{\rm s}}
\newcommand{\uT}{\,\mu{\rm T}}

\newcommand{\ye}{\gamma_\mr{e}}
\newcommand{\yn}{\gamma_\mr{n}}
\newcommand{\angstrom}{\textup{\AA}}

\newcommand{\ms}{m_S}

\newcommand{\aiso}{a_\mr{iso}}
\newcommand{\apar}{a_{||}}
\newcommand{\aperp}{a_{\perp}}

\newcommand{\Bo}{B_0}

\newcommand{\rDFT}{r_\mr{DFT}}

\newcommand{\thetaDFT}{\theta_\mr{DFT}}

\newcommand{\vecA}{\vec{A}_z}
\newcommand{\vecB}{\vec{B}_0}
\newcommand{\vecDB}{\Delta\vec{B}}
\newcommand{\vecBtot}{\vec{B}_\mr{tot}}
\newcommand{\vecr}{\vec{r}}

\newcommand{\vecz}{\vec{e}_z}
\newcommand{\vecS}{\vec{S}}

 

\begin{document}

\title{Three-dimensional localization spectroscopy of individual nuclear spins with sub-Angstrom resolution}
\author{J. Zopes$^1$, K. S. Cujia$^1$, K. Sasaki$^{1,2}$, J. M. Boss$^1$, K. M. Itoh$^2$, and C. L. Degen$^{1,}$\footnote{Email: degenc@ethz.ch}}
\affiliation{$^1$Department of Physics, ETH Zurich, Otto Stern Weg 1, 8093 Zurich, Switzerland.}
\affiliation{$^2$School of Fundamental Science and Technology, Keio University, Yokohama 223-8522, Japan.}


\begin{abstract}
We report on precise localization spectroscopy experiments of individual \C nuclear spins near a central electronic sensor spin in a diamond chip. By detecting the nuclear free precession signals in rapidly switchable external magnetic fields, we retrieve the three-dimensional spatial coordinates of the nuclear spins with sub-Angstrom resolution and for distances beyond $10\unit{\angstrom}$.  We further show that the Fermi contact contribution can be constrained by measuring the nuclear $g$-factor enhancement.  The presented method will be useful for mapping the atomic-scale structure of single molecules, an ambitious yet important goal of nanoscale nuclear magnetic resonance spectroscopy.
\end{abstract}


\maketitle
One of the visionary goals of nanoscale quantum metrology with nitrogen-vacancy (NV) centers is the structural imaging of individual molecules, for example proteins, that are attached to the surface of a diamond chip \cite{shi15}. By adapting and extending measurement techniques from nuclear magnetic resonance (NMR) spectroscopy, the long-term perspective is to reconstruct the chemical species and three-dimensional location of the constituent atoms with sub-Angstrom resolution \cite{ajoy15,perunicic16}. In contrast to established structural imaging techniques like X-ray crystallography, cryo-electron tomography or conventional NMR, which average over large numbers of target molecules, only a single copy of a molecule is required.  Conformational differences between individual molecules could thus be directly obtained, possibly bringing new insights about their structure and function.

In recent years, first experiments that address the spatial mapping of nuclear and electron spins with NV based quantum sensors have been devised.  One possibility is to map the position into a spectrum, as it is done in magnetic resonance imaging.  For nanometer-scale imaging, this requires introducing a nanomagnet \cite{degen09,grinolds14,lazariev15}. Another approach is to exploit the magnetic gradient of the NV center's electron spin itself, whose dipole field shifts the resonances of nearby nuclear spins as a function of distance and internuclear angle.  Refinements in quantum spectroscopic techniques have allowed the detection of up to 8 individual nuclear spins \cite{taminiau12,schirhagl14} as well as of spin pairs \cite{shi14,reiserer16,abobeih18} for distances of up to $\sim 30\unit{\angstrom}$ \cite{zhao12,maurer12}.
Due to the azimuthal symmetry of the dipolar interaction, however, these measurements can only reveal the radial distance $r$ and polar angle $\theta$ of the inter-spin vector $\vecr = (r,\theta,\phi)$, but are unable to provide the azimuth $\phi$ required for reconstructing three-dimensional nuclear coordinates.  One possibility for retrieving $\phi$ is to change the direction of the static external field \cite{zhao12}, however, this method leads to a mixing of the NV center's spin levels which suppresses the ODMR signal \cite{epstein05} and shortens the coherence time \cite{stanwix10}.  Other proposed methods include position-dependent polarization transfer \cite{laraoui15} or combinations of microwave and radio-frequency fields \cite{wang16,sasaki18}.
%
\begin{figure}[h!]
\includegraphics[width=0.48\textwidth]{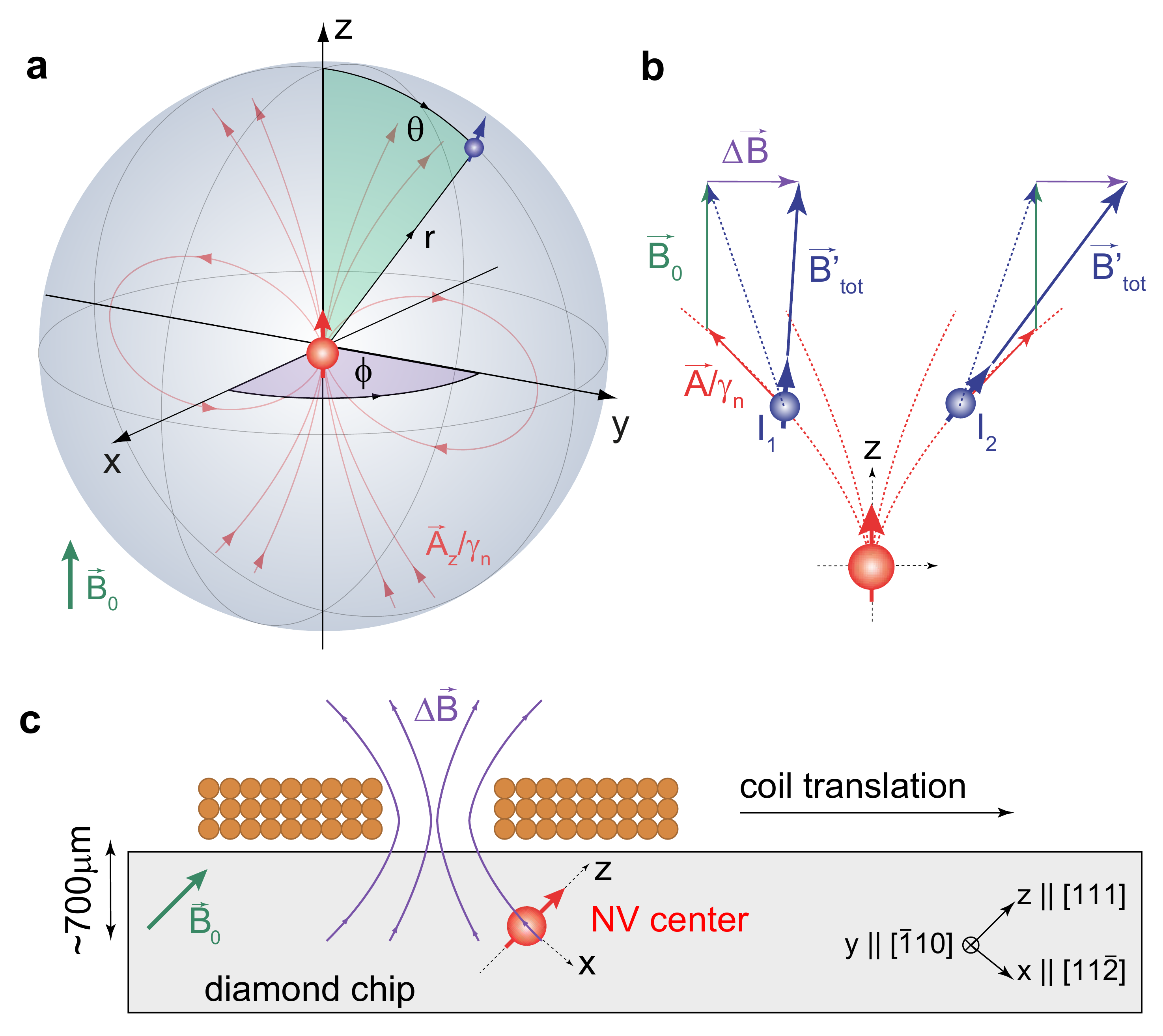}
\caption{\captionstyle
{\bf Coordinate systems for spins and magnetic fields.}
({\bf a}) Reference frame of the central nitrogen-vacancy (NV) sensor spin (red) with a nuclear spin (blue) located at the three-dimensional position $\vecr = (r,\theta,\phi)$. The quantization axis of the NV center defines the $z$-axis.  The hyperfine field of the NV spin (red field lines) provides the magnetic field gradient for imaging.
({\bf b}) Sketch of two nuclear spins $I_1$ and $I_2$ experiencing the same hyperfine interaction (red) [Eq. (\ref{eq:hyperfinevector})].  Application of a transverse field $\vecDB$ (purple) reduces ($I_1$) or increases ($I_2$) the total magnetic field $\vecBtot'$ (blue) experienced by the nuclear spins depending on the $\phi$ angle, allowing us to discriminate the nuclear locations.  $\Bo$ is the static external field (green).
({\bf c}) Geometry of the experimental setup in the laboratory frame of reference.  A small solenoid on top of the diamond chip provides a rapidly switchable magnetic field $\vecDB$.  To change the vector orientation of $\vecDB$, we translate the coil over the diamond.}
\label{fig:fig1}
\end{figure}

Here, we demonstrate three-dimensional localization of individual, distant nuclear spins with sub-Angstrom resolution.  To retrieve the ``missing angle'' $\phi$, we combine a dynamic tilt of the quantization axes using a high-bandwidth microcoil with high resolution correlation spectroscopy \cite{laraoui13,boss16}. Our method provides the advantage that manipulation and optical readout of the electronic spin can be carried out in an aligned external bias field. This ensures best performance of the optical readout and the highest magnetic field sensitivity and spectral resolution of the sensor.


We consider a nuclear spin $I=1/2$ located in the vicinity of a central electronic spin $S=1$ with two isolated spin projections $\ms = \{0,-1\}$.  The nuclear spin experiences two types of magnetic field, a homogeneous external bias field $\Bo$ (aligned with the quantization axis $\vecz$ of the electronic spin), and the local dipole field of the electronic spin. Because the electronic spin precesses at a much higher frequency than the nuclear spin, the latter only feels the static component of the electronic field, and we can use the secular approximation to obtain the nuclear free precession frequencies,
\begin{align}
f_{\ms} = \frac{1}{2\pi} || -\yn \vecBtot || = \frac{1}{2\pi} || - \yn\Bo\vecz + \ms \vecA(\vecr)|| \ .
\label{eq:precessionfrequencies}
\end{align}
Here, $\yn$ is the nuclear gyromagnetic ratio and
\begin{align}
\vecA(\vecr) &= \textbf{A}(\vecr)\cdot\vec{e}_z =(A_{xz},A_{yz},A_{zz})  \nonumber \\
 &= ( \aperp\cos(\phi),\aperp\sin(\phi),\apar ) 
\label{eq:hyperfinevector}
\end{align}
is the secular hyperfine vector of the hyperfine tensor $\textbf{A}(\vecr)$ that gives rise to the hyperfine magnetic field $\ms\vecA(\vecr)/\yn$ (see Fig. \ref{fig:fig1}b).

To obtain information about the distance vector $\vecr$, a standard approach is to measure the parallel and transverse components of the hyperfine vector, $\apar = A_{zz}$ and $\aperp = (A_{xz}^2+A_{yz}^2)^{1/2}$, and to relate them to the field of a point dipole,
\begin{align}
\apar &= \frac{\mu_0\ye\yn\hbar}{4\pi r^3} \, (3\cos^2\theta-1) + \aiso \ , \label{eq:apar} \\
\aperp &= \frac{\mu_0\ye\yn\hbar}{4\pi r^3} \, 3\sin\theta\cos\theta \ , \label{eq:aperp} 
\end{align}
where $\mu_0 = 4\pi \cdot 10^{-7}\unit{T\cdot m/A}$ is the vacuum permeability, $\hbar=1.054\ee{-34}\unit{J\cdot s}$ is the reduced Planck constant, $|\ye |= 2\pi \cdot 28\unit{GHz/T}$ is the electron gyromagnetic ratio,
and where we have included a Fermi contact term $\aiso$ (set to zero for now) for later discussion.
Experimentally, the parallel projection $\apar$ can be inferred from the precession frequencies $f_{\ms}$ using Eq. (\ref{eq:precessionfrequencies}), and the transverse projection $\aperp$ can be determined by driving a nuclear Rabi rotation via the hyperfine field of the central spin and measuring the rotation frequency \cite{boss16}.  Once $\apar$ and $\aperp$ are known, Eqs. (\ref{eq:apar},\ref{eq:aperp}) can be used to extract the distance $r$ and polar angle $\theta$ of the distance vector $\vecr=(r,\theta,\phi)$.  Due to the rotational symmetry of the hyperfine interaction, however, knowledge of $\apar$ and $\aperp$ is insufficient for determining the azimuth $\phi$.
%
\begin{figure}[t!]
\includegraphics[width=0.48\textwidth]{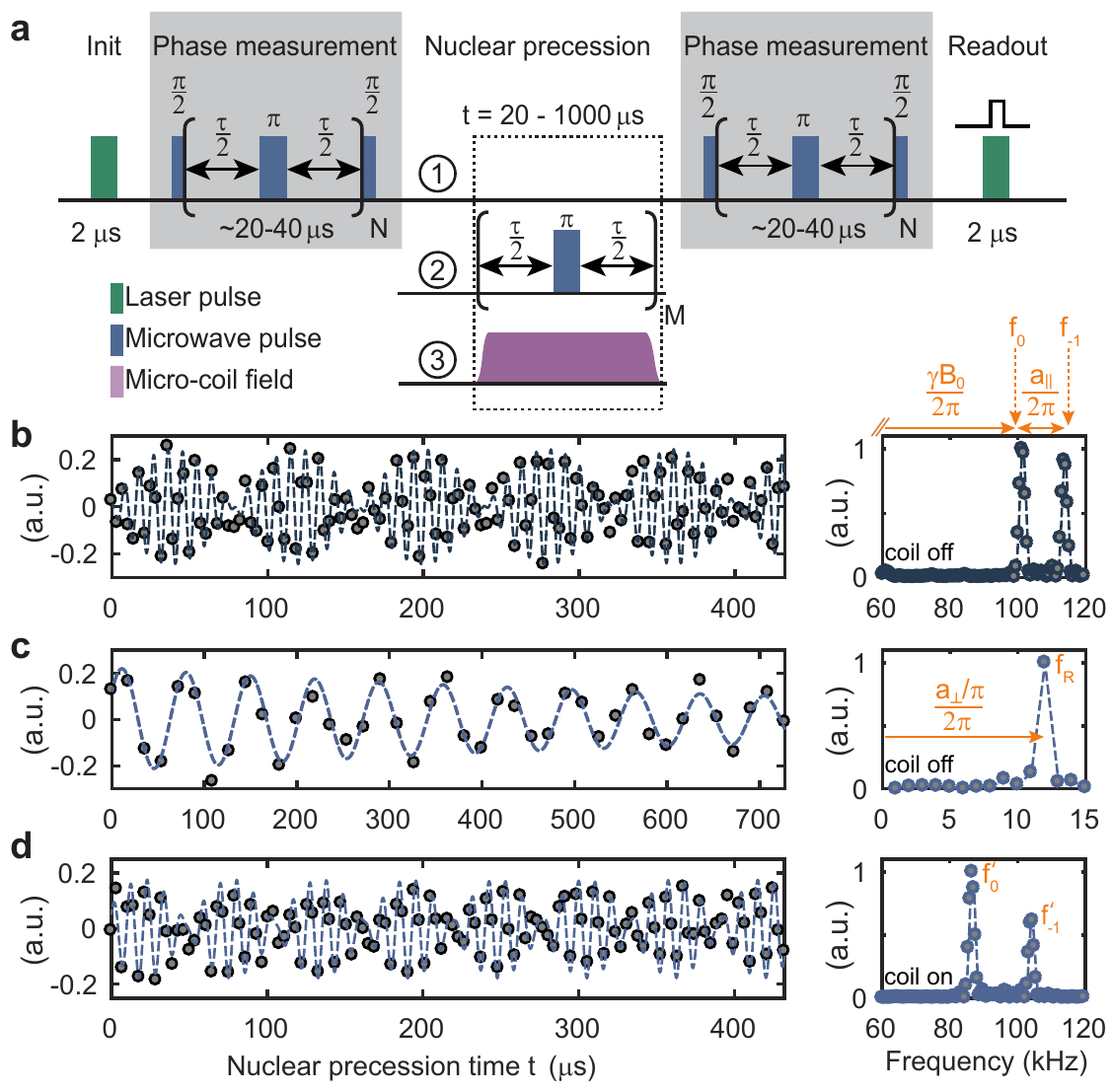} 
\caption{\captionstyle
{\bf Implementation of three-dimensional localization spectroscopy.}
({\bf a}) Correlation spectroscopy protocol.  By correlating two phase measurements we trace out the precession of the target nuclear spin(s) under different NMR sequences. Phase measurements are implemented by a Carr-Purcell-Meiboom-Gill (CPMG) train of microwave $\pi$ pulses (blue) applied to the central electronic spin, where $\tau\approx[2(f_0+f_{-1})]^{-1}$.  Laser pulses (green) are used to polarize and read out the electronic spin.  Repetitions are $N=4-8$ (see Ref. \cite{supplementary}) and $M=t/\tau$.
({\bf b}) Free precession signal of the nuclear spin as a function of time $t$, using sequence \ding{172}.  Right panel shows the corresponding power spectrum.  The two frequencies $f_0$ and $f_{-1}$ are approximately equal to $\yn\Bo/(2\pi)$ and $(\yn\Bo+\apar)/(2\pi)$, respectively, see text.
({\bf c}) Application of periodic $\pi$ pulses on the NV center during $t$ (sequence \ding{173}) causes a Rabi nutation of the nuclear spin, whose oscillation frequency $f_R$ is approximately equal to $(\aperp/\pi)/(2\pi)$.
({\bf d})	Activation of a transverse microcoil field $\vecDB$ during the nuclear precession (sequence \ding{174}) leads to shifted frequencies $f_0'$ and $f_{-1}'$.
All measurements were conducted on \Cn{1}. Extracted frequencies are listed in Table \ref{tab:table1}.
}
\label{fig:fig2}
\end{figure}

To break the rotational symmetry and recover $\phi$, we apply a small transverse magnetic field $\vecDB$ during the free precession of the nuclear spin. Application of a transverse field tilts the quantization axes of the nuclear and electronic spins.  The tilting modifies the hyperfine coupling parameters $\apar$ and $\aperp$ depending on the angle between $\vecDB$ and $\vecA$, which in turn shifts the nuclear precession frequencies $f_{\ms}$.  To second order in perturbation theory, the $\ms$-dependent precession frequencies are given by \cite{childress06}
\begin{align}
f_{\ms} &= \frac{1}{2\pi} ||-\yn\vecBtot'||  \\
  &= \frac{1}{2\pi} ||-\yn\Bo\vecz -\yn(1+\alpha(\ms)) \vecDB + \ms\vecA(\vecr)||,
\label{eq:precfreq}
\end{align}
where $\alpha(\ms)$ is a small enhancement of the nuclear $g$-factor.  The enhancement results from non-secular terms in the Hamiltonian that arise due to the tilting of the electronic quantization axis, and is given by \cite{childress06}
\begin{equation}
\alpha(\ms) \approx (3|\ms|-2) \frac{\gamma_e}{\gamma_n D}   \begin{pmatrix}
    A_{xx} & A_{xy} & A_{xz} \\
    A_{yx} & A_{yy} & A_{yz} \\
    0	      & 0          & 0 
\end{pmatrix} \ .
\label{eq:gfactor}
\end{equation}
Here $D =2\pi \times 2.87\,\mathrm{GHz}$ is the ground-state zero-field splitting of the NV center.  By measuring the shifted frequencies $f_{\ms}$ and comparing them to the theoretical model of Eqs. (\ref{eq:precfreq},\ref{eq:gfactor}), we can then determine the relative $\phi$ angle between the hyperfine vector and $\vecDB$.



We experimentally demonstrate three-dimensional localization spectroscopy of four $^{13}\mathrm{C}_{1-4}$ nuclei adjacent to three distinct NV centers.  NV$_1$ is coupled to two \C spins, while NV$_2$ and NV$_3$ are each coupled to a single \C spin.  For read-out and control of the NV center spin, we use a custom-built confocal microscope that includes a coplanar waveguide and a cylindrical permanent magnet for providing an external bias field of $\Bo \sim 10\unit{mT}$ applied along the NV center axis $\vecz$.  Precise alignment of the bias field is crucial for our experiments and is better than $0.3^\circ$ \cite{supplementary}.

To dynamically tilt the external field we implement a multi-turn solenoid above the diamond surface (see Fig. \ref{fig:fig1}d).  The coil produces $\sim 2.5\unit{mT}$ field for $600 \unit{mA}$ of applied current and has a rise time of $\sim 2\unit{\us}$.  We calibrate the vector magnetic field of the coil with an absolute uncertainty of less than $15\unit{\uT}$ in all three spatial components using two other nearby NV centers with different crystallographic orientations \cite{steinert10,supplementary}.
%
\begin{table}[b!]
    \begin{tabular}{l|l|l}
		\hline\hline
      Quantity  & Value                     & Reference									\\ \hline
     $ f_0, f_{-1}   $       &  $101.7(1),114.2(1) \,\mathrm{kHz}$  & Fig. \ref{fig:fig2}b \\
     $f_R$                     & $14.4(1)\,\mathrm{kHz}$      & Fig. \ref{fig:fig2}c                   \\ \hline
     $f'_0$, $f'_{-1}$     & $88.3(3),103.2(2)\,\mathrm{kHz}$      & Fig. \ref{fig:fig2}d  \\ \hline 
 $\vec{B}_{0}$                & $(0.028,-0.056,9.502)\,\mathrm{mT}$    & Ref. \cite{supplementary}                    \\
        $\Delta \vec{B}$ & $(-1.715,  0.614,  -1.547)\,\mathrm{mT}$ & Ref. \cite{supplementary}                  \\ 
        \hline \hline
    \end{tabular}
\caption{Data base of measured precession frequencies and calibrated external magnetic fields used to determine the 3D position of \Cn{1}.
Five further measurements of $(f'_0, f'_{-1})$ were made to improve the localization accuracy (data given in Ref. \cite{supplementary}).
Vector magnetic fields refer to the NV coordinate system defined in Fig. \ref{fig:fig1}.}
\label{tab:table1}
\end{table}


We begin our 3D mapping procedure by measuring the parallel and perpendicular hyperfine coupling constants using conventional correlation spectroscopy \cite{boss16} with no coil field applied, $\vecDB=0$ (Fig. \ref{fig:fig2}).  The parallel coupling $\apar$ is determined from a free precession experiment (sequence \ding{172} in Fig. \ref{fig:fig2}) yielding the frequencies $f_0$ and $f_{-1}$ (Fig. \ref{fig:fig2}b).  The coupling constant is then approximately given by $\apar/(2\pi) \approx f_{-1} - f_0$.  The transverse coupling $\aperp$ is obtained by driving a nuclear Rabi oscillation via the NV spin, using sequence \ding{173}, and recording the oscillation frequency $f_R$, where $\aperp/(2\pi) \approx \pi f_R$ (Fig. \ref{fig:fig2}c).  Because the Zeeman and hyperfine couplings are of similar magnitude, these relations are not exact and proper transformation must be applied to retrieve the exact coupling constants $\apar$ and $\aperp$ \cite{boss16,supplementary}.  Once the hyperfine parameters are known, we can calculate the radial distance $r=8.58(1)\unit{\angstrom}$ and the polar angle $\theta=52.8(1)^\circ$ of the nuclear spin by inverting the point-dipole formulas (\ref{eq:apar},\ref{eq:aperp}).  The measurement uncertainties in $r$ and $\theta$ are very small because correlation spectroscopy provides high precision estimates of both $\apar$ and $\aperp$.


In a second step, we repeat the free precession measurement with the coil field turned on (sequence \ding{174}), yielding a new pair of frequency values $f_0'$, $f_{-1}'$ (Fig. \ref{fig:fig2}d).  We then retrieve $\phi$ by computing theoretical values for $f_{0}^\mr{(th)}$, $f_{-1}^\mr{(th)}$ based on Eq. (\ref{eq:precfreq}) and the calibrated fields in Table \ref{tab:table1}, and minimizing the cost function
\begin{align}
\xi(\phi) = [f'_{-1}-f'_{0}] - [f_{-1}^\mr{(th)}(\phi)-f_{0}^\mr{(th)}(\phi)] \ .
\label{eq:xi}
\end{align}
with respect to $\phi$.  To cancel residual shifts in the static magnetic field and improve the precision of the estimates, we compare the frequency difference between $\ms$ states rather than the absolute precession frequencies.
%
\begin{figure}[t!]
\includegraphics[width=0.48\textwidth]{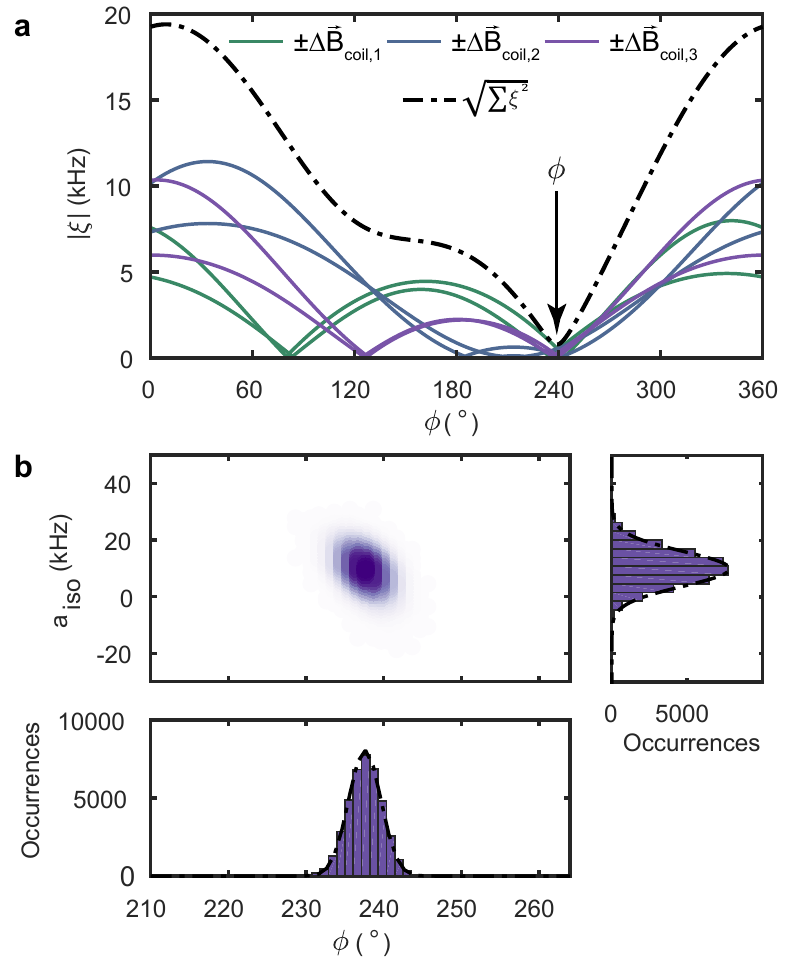}
\caption{\captionstyle
{\bf Determination of azimuth angle $\phi$ and Fermi contact contribution $\aiso$ for \Cn{1}.}
({\bf a}) Cost function $|\xi (\phi)|$ between observed and predicted precession frequencies, as defined in Eq. (\ref{eq:xi}).  Here $\aiso=0$.
Six measurements are shown for three spatial coil positions (solid curves) and opposite polarities of the coil current.
The estimate for $\phi$ is given by the minimum of the squared cost functions $\sum {|\xi (\phi)|^2}$ of the six measurements (dash-dotted curve).
({\bf b}) Scatter plot of maximum likelihood estimates of $\phi$ and $\aiso$ obtained by Monte Carlo error propagation. The plot is generated from $4\cdot 10^4$ scatter points,
where each point is the result of minimizing $\sum {|\xi (\phi,\aiso)|^2}$ for a different Monte Carlo sampling.
Histograms for $\phi$ (bottom) and $\aiso$ (right) are obtained by integrating the 2D scatter plot along the vertical or horizontal direction, respectively.
Corresponding plots for \Cn{2-4} are given in Ref. \cite{supplementary}.
\label{fig:fig3}
}
\end{figure}
%
\begin{table*}[t!]
\centering
\begin{ruledtabular}
\begin{tabular}{c|ccc|ccc|c|ccc|cc}
\ & \multicolumn{6}{c|}{Experimental values} & \multicolumn{6}{c}{DFT values \cite{nizovtsev18}} \\
Atom & $\apar$/kHz & $\aperp$/kHz & $\aiso$/kHz  & $r$/\angstrom & $\theta/^\circ$ & $\phi/^\circ$ &
Lattice sites$^a$    & $\apar$/kHz & $\aperp$/kHz & $\aiso$/kHz  & $\rDFT$/\angstrom & $\thetaDFT/^\circ$ \\
\hline
\Cn{1} &   3.1(1) & 44.5(1) &  9(8)   &  8.3(2)  & 58(4)   & 238(2) & \{386,395,447\} &   1.3 & 43.2 &  4.0 &  8.6     & 60$^b$ \\
\Cn{2} & 119.0(1) & 65.9(1) & 19(15)  &  6.8(3)  & 19(3)   &  20(5) & \{33,39,41\}    & 100.4 & 64.8 & -2.4 &  6.3     & 24$^b$ \\
\Cn{3} &  18.5(1) & 41.4(2) &  1(6)   &  8.9(1)  & 43(4)   & 208(4) & \{450,455,466\} &  15.9 & 37.8 &  1.7 &  9.2     & 45$^b$ \\
\Cn{4} &   1.9(1) & 19.2(1) & ---$^c$ & 11.47(1) & 51.8(2) &  34(4) & \multicolumn{6}{c}{---$^d$} \\
\end{tabular}
\end{ruledtabular}
\caption{Measured hyperfine couplings and inferred 3D locations of \C nuclei measured on three NV centers.
Errors are one standard deviation and represent the confidence interval from the Monte Carlo error propagation according to Fig. \ref{fig:fig3}b.
DFT values are for the lattice site(s) whose calculated hyperfine couplings best match the experimental data.
$^a$Ref. \cite{nizovtsev18} does not specify the $\phi$ angle, therefore, three symmetric sites are compatible with our data.
$^b$Due to the inversion symmetry of the hyperfine interaction, our method cannot distinguish between sites in the upper and lower hemisphere;  the table therefore lists $\min(\thetaDFT,180^\circ-\thetaDFT)$.
$^c$Constrained to $\aiso=0$.
$^d$No DFT data available. }
\label{tab:table2}
\end{table*}

In Fig. \ref{fig:fig3}a, we plot $|\xi(\phi)|$ for three different coil positions and opposite coil currents for \Cn{1}.  We use several coil positions because a single measurement has two symmetric solutions for $\phi$, and also because several measurements improve the overall accuracy of the method.  The best estimate $\phi = 239(2)^\circ$ is then given by the least squares minimum of the cost functions (dash-dotted line in Fig. \ref{fig:fig3}a). To obtain a confidence interval for $\phi$, we calculate a statistical uncertainty for each measurement by Monte Carlo error propagation taking the calibration uncertainties in $\vecB$ and $\vecDB$, as well as the measurement uncertainties in the observed precession frequencies into account \cite{supplementary}. Values for all investigated \C nuclei are collected in Ref. \cite{supplementary}.


Thus far we have assumed that the central electronic spin generates the field of a perfect point dipole.  Previous experimental work \cite{smeltzer11,childress06} and density functional theory (DFT) simulations \cite{gali08,nizovtsev18}, however, suggest that the electronic wave function extends several Angstrom into the diamond host lattice.  The finite extent of the spin density leads to two deviations from the point dipole model: (i) modified hyperfine coupling constants $A_{ij}$, and (ii) a non-zero Fermi contact term $\aiso$.  In the remainder of this study we estimate the systematic uncertainty to the localization of the nuclear spins due to deviations from the point dipole model.

We first consider the influence of the Fermi contact interaction, which arises from a non-vanishing NV spin density at the location of the nuclear spin.  The Fermi contact interaction adds an isotropic term to the hyperfine coupling tensor, $\textbf{A}+\aiso\textbf{1}$, which modifies the diagonal elements $A_{xx}$, $A_{yy}$ and $A_{zz}$.  DFT simulations \cite{gali08,nizovtsev18} indicate that $\aiso$ can exceed $100\unit{kHz}$ even for nuclear spins beyond $7\unit{\angstrom}$.  It is therefore important to experimentally constrain the size of $\aiso$.

To determine $\aiso$, one might consider measuring the contact contribution to the parallel hyperfine parameter $\apar$, which is equal to $A_{zz}$.  This approach, however, fails because a measurement of $\apar$ cannot distinguish between dipolar and contact contributions.  Instead, we here exploit the fact that the gyromagnetic ratio enhancement $\alpha$ depends on $A_{xx}$ and $A_{yy}$, and hence $\aiso$.
To quantify the Fermi contact coupling we include $\aiso$ as an additional free parameter in the cost function (\ref{eq:xi}). By minimizing $\xi(\phi,\aiso)$ as a joint function of $\phi$ and $\aiso$ and generating a scatter density using Monte Carlo error propagation, we obtain maximum likelihood estimates and confidence intervals for both parameters (Fig. \ref{fig:fig3}b).  The resulting contact coupling and azimuth for nuclear spin \Cn{1} are $\aiso/(2\pi) = 9(8)\unit{kHz}$ and $\phi = 238(2)^\circ$, respectively; data for \Cn{2-4} are collected in Table \ref{tab:table2}.  Because the gyromagnetic ratio enhancement $\alpha$ is only a second-order effect, our estimate is poor, but it still allows us constraining the size of $\aiso$.  By subtracting the Fermi contact contribution from $\apar$, we further obtain refined values for the radial distance and polar angle, $r=8.3(2)\unit{\angstrom}$ and $\theta=58(4)^\circ$.
Note that introducing $\aiso$ as a free parameter increases the uncertainties in the refined $r$ and $\theta$, because the error in $\aiso$ is large.  This leads to disproportionate errors for distant nuclei where $\aiso$ is small.  Once nuclei are beyond a certain threshold distance, which we set to $r=10\unit{\angstrom}$ in Table \ref{tab:table2}, it therefore becomes more accurate to constrain $\aiso=0$ and apply the simple point dipole model.

\begin{figure}[b!]
\includegraphics[width=0.48\textwidth]{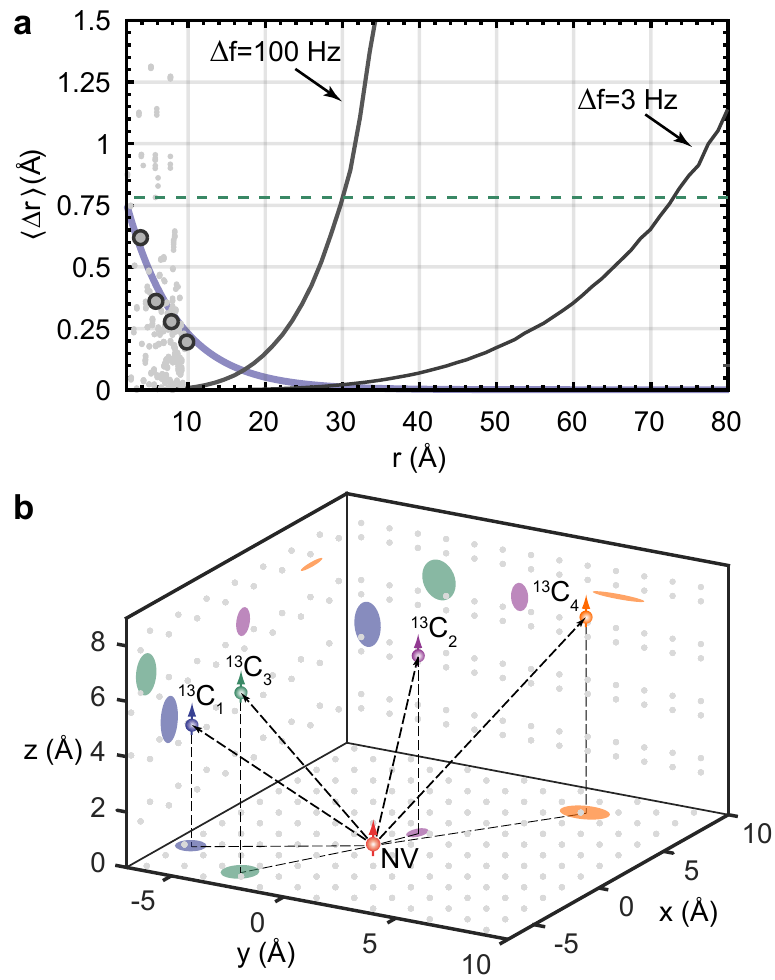}
\caption{\captionstyle
{\bf Three-dimensional localization of four \C nuclear spins.}
({\bf a}) Average localization uncertainty $\langle\Delta r\rangle$ as a function of radial distance $r$ to the central spin.  Gray dots represent the systematic error of the point-dipole approximation (see text), extracted for all lattice sites reported in the DFT calculation of Ref. \cite{nizovtsev18} .  Blue curve is an exponential fit to the median values (gray circles) of the gray dots in intervals of $2\unit{\angstrom}$.  Black curves show the uncertainty of the frequency measurement, assuming a precision of $100\unit{Hz}$ (this study) and of $3\unit{Hz}$ \cite{rosskopf17,zaiser16,abobeih18}.  Dashed horizontal line is one-half the diamond C-C bond length.
({\bf b}) Reconstructed locations of the four distant nuclear spins \Cn{1-4}. Shaded regions mark the $2\sigma$-confidence area of the localization projected onto ($xy$,$yz$,$xz$)-planes of the coordinate system. Gray points represent carbon lattice positions projected onto the same planes.  The origin is set to the expected center of gravity of the spin density at $2.29\unit{\angstrom}$ from the nitrogen nucleus on the N-V symmetry axis \cite{gali08,nizovtsev18}.  Due to the inversion symmetry of the hyperfine interaction, our method cannot distinguish between sites in the upper and lower hemisphere; all \C are therefore plotted in the upper hemisphere.
\label{fig:fig4}
}
\end{figure}

The second systematic error in the position estimate results from the finite size of the NV center's electronic wave function.  Once the extent of the wave function becomes comparable to $\vecr$,  the anisotropic hyperfine coupling constants $A_{ij}$ are no longer described by a point dipole, but require integrating a geometric factor over the sensor spin density \cite{gali08}.  While we cannot capture this effect experimentally, we can estimate the localization uncertainty from DFT simulations of the NV electron spin density.  Following Ref. \cite{nizovtsev18}, we convert the calculated DFT hyperfine parameters of 510 individual lattice sites to $(r,\theta)$ positions using the point-dipole formula (\ref{eq:apar},\ref{eq:aperp}), and compute the difference to the DFT input parameters $(\rDFT,\thetaDFT)$.  The result is plotted in Fig. \ref{fig:fig4}a.  We find that the difference $\langle\Delta r\rangle = r-\rDFT$ decreases roughly exponentially with distance, and falls below $0.2\unit{\angstrom}$ when $r>10\unit{\angstrom}$ (grey dots and curve).
%


Fig. \ref{fig:fig4}b summarizes our study by plotting the reconstructed locations for all four carbon atoms in a combined 3D chart.  The shaded regions represent the confidence areas of the localization, according to Table \ref{tab:table2}, projected onto the Cartesian coordinate planes.
We note that the DFT simulations are in good agreement with our experimental results.
The accuracy of our present experiments is limited by deviations from the point-dipole model, which dominate for small $r$ (see Fig. \ref{fig:fig4}a).  In future experiments that probe more distant nuclear spins, this systematic uncertainty will be much smaller, and the localization precision will eventually be dictated by the frequency resolution of our nuclear precession measurement.  While the frequency precision was of order $100\unit{Hz}$ in the present study, much work has recently been put into improving the frequency resolution \cite{rosskopf17,boss17,schmitt17}.  Assuming a precision of $3\unit{Hz}$ \cite{glenn18,abobeih18}, the projected radial uncertainty at $70\unit{\angstrom}$ is below $0.7\unit{\angstrom}$, which is less than one-half the C-C bond length of $1.54\unit{\angstrom}$ (see Fig. \ref{fig:fig4}a).  Such a precision is in principle sufficient to analyze the interior structure of single molecules deposited on a diamond chip, assuming adequate detection sensitivity.

To conclude, we have demonstrated precise localization of four \C nuclear spins with sub-Angstrom resolution in all three spatial dimensions, and for radial distances exceeding $10\unit{\angstrom}$.  By analyzing the $g$-factor enhancement in an off-axis magnetic field, we were further able to constrain the Fermi contact contribution.  Looking forward, our technique can be extended by measuring nuclear spin-spin interactions \cite{shi14,reiserer16,abobeih18}, which will provide important structural constraints for molecular modeling.  In addition, our strategy can be combined with methods for signal enhancement, like as nanostructured sensor chips \cite{wan17} or hyperpolarization techniques \cite{abrams14}.  All of these advances will be critical for realizing the long term goal of imaging of single molecules with atomic resolution, which will have many applications in structural biology and chemical analytics.


{\bf Acknowledgments:}
The authors thank Jyh-Pin Chou and Adam Gali for sharing DFT data, and Julien Armijo, Kevin Chang, Nils Hauff, Konstantin Herb, Takuya Segawa and Tim Taminiau for helpful discussions.  This work was supported by Swiss NSF Project Grant $200021\_137520$, the NCCR QSIT, and the DIADEMS programme 611143 of the European Commission.  The work at Keio has been supported by JSPS KAKENHI (S) No. 26220602, JSPS Core-to-Core, and Spin-RNJ.





{\bf Methods:}

\textit{Diamond sample:}
Experiments were performed on a bulk, electronic-grade diamond crystal from ElementSix with dimensions $2\,\mathrm{mm} \times 2\,\mathrm{mm} \times 0.5\,\mathrm{mm}$ with \hkl<110> edges and a \hkl<100> front facet. The diamond was overgrown with a layer structure of $20\,\mathrm{nm}$ enriched $^{12}\mathrm{C}$ (99.99\,\%), $1\,\mathrm{nm}$ enriched $^{13}\mathrm{C}$ (estimated in-grown concentration $\sim 5-10\,\%$) and a $5\,\mathrm{nm}$ cap layer of again enriched $^{12}\mathrm{C}$ (99.99\,\%). Nitrogen-vacancy (NV) centers were generated by ion-implantation of \NN with an energy of $5\,\mathrm{keV}$, corresponding to a depth of $\sim 5-10\,\mathrm{nm}$. After annealing the sample for NV formation, we had to slightly etch the surface (at $580 ^{\circ}\mathrm{C}$ in pure $\mathrm{O}_2$) to remove persistent surface fluorescence. The intrinsic nuclear spin of the three NV centers studied in our experiments were confirmed to be of the \NN isotope. Further characterizations and details on the sample can be found in a recent study (sample B in Ref. \cite{unden18}).

\textit{Coordinate systems:} 
In Supplementary Fig. S2a both laboratory and NV coordinate system are shown in a combined schematic. The laboratory coordinate system ($x_{\mathrm{Lab}}$,$y_{\mathrm{Lab}}$,$z_{\mathrm{Lab}}$) is defined by the normal vectors to the diamond faces, which lie along \hkl<110>,\hkl<-110> and \hkl<001>, respectively. The reference coordinate system of the NV center is defined by its quantization direction, which is labelled $z_{\mathrm{NV}}$ and lies along \hkl<111>. The $x_{\mathrm{NV}}$- and  $y_{\mathrm{NV}}$-axis are pointing along the \hkl<11-2> and \hkl<-110> direction, respectively. 

\textit{Experimental apparatus:}
A schematic of the central part of the experimental apparatus is shown in Supplementary Fig. S1. The diamond sample is glued to a $200\,\mu\mathrm{m}$ thick glass piece and thereby held above a quartz slide with incorporated microwave transmission line for electron spin control. Below the quartz slide we placed a high numerical aperture (NA= 0.95)
microscope objective for NV excitation with a $532\,\mathrm{nm}$ laser and detection using a single photon counting module (SPCM). 
We applied static, external magnetic bias fields with a cylindrical NdFeB permanent magnet (not shown in Supplementary Fig. S1). The magnet is attached to a motorized, three-axis translation stage. The NV control pulses  were  generated  by  an  arbitrary  waveform  generator  (Tektronix,  AWG5002C) and upconverted by I/Q mixing with a local oscillator to the desired $\sim 2.6\,\mathrm{GHz}$. 

\textit{Planar, high-bandwidth coil:}
The planar coil is positioned directly above the diamond sample and attached to a metallic holder, which can be laterally shifted to translate the coil. Due to the thickness of the diamond ($500\,\mu\mathrm{m}$) and the glass slide the minimal vertical stand-off of the coil to the NV centers is approximately $700\,\mu\mathrm{m}$. Design parameters of the planar coil, used in our experiments, are listed in Supplementary Table S1. These were found by numerically maximizing the magnetic field at the position of the NV center, located at a planned vertical stand-off of $\sim 700-1000\,\mu\mathrm{m}$ (see Supplementary Fig. S1).  The coil had an inductance of $\le 2.5\,\mu\mathrm{H}$ and a resistance of $\le 0.5\,\Omega$.  The coil was manufactured by Sibatron AG (Switzerland) and it is mounted onto a copper plate, that acts as a heat-sink, using thermally conducting glue. For the coil control, a National Instruments NI PCI 5421 arbitrary waveform generator was used, to generate voltage signals that controlled a waveform amplifier (Accel Instruments TS-250) which drives the coil current.

\textit{Calibration of the coil field $\vecDB$:}
We calibrated the vector field generated by the coil $\vecDB$ using the target NV, coupled to nuclear spins of interest, and two auxiliary NV centers with different crystallographic orientation. All three NV centers were located in close proximity to each other, with a distance of typically $\le 5\,\mu\mathrm{m}$ (see Supplementary Fig. S2c). Over this separation the magnetic field of the coil can be assumed to be homogeneous. We determined the orientation of the symmetry axis of many NV centers by moving the permanent magnet over the sample and observing the ODMR splitting. The azimuthal orientation of the target NVs defines the x-axis in the laboratory and NV frame ($\phi=0$). This orientation was the same for all target NV centers investigated in this work. The auxiliary NV centers were selected to be oriented along $\phi_{a_1}=90^{\circ}$ and $\phi_{a_2}=270^{\circ}$ (see Supplementary Fig. S2b). To calibrate the coil field, we removed the permanent magnet and recorded ODMR spectra for the target NV center and both auxiliary NV centers with the field of the coil activated. In this way we record in total 6 ODMR lines, with 2 lines per NV center.

A numerical, nonlinear optimization method was used to determine the magnetic field $\vecDB$ from these ODMR resonances. For each of the three NV centers we simultaneously minimized the difference between the measured ODMR lines and the eigenvalues of the ground-state Hamiltonian:
\begin{align}
H_i = D S_z^2 + \ye (\vecDB)_i \cdot\vecS.
\end{align}
Here, the magnetic field $(\vecDB)_i$ acting onto the specific NV center is obtained by a proper rotation of $\vecDB$ into the respective reference frame. 

\textit{Precise alignment of the bias field $\vecB$:}
Precise alignment of the external bias field to the quantization axis of the NV center (z-axis) is critical for azimuthal localization measurements, because residual transverse fields of $\vecB$ modify the precession frequencies in the same way as the field of the coil. The coarse alignment of the magnet and a rough adjustment of the magnitude of the field, to $\sim 10\, \mathrm{mT}$, was achieved by recording ODMR spectra of the target NV center for different $(x,y,z)$-positions of the magnet.  Afterwards, we iteratively optimized the alignment of the magnet. In each iteration, we reconstructed the vector field $\vecB$ acting on the target NV centers using the method used for the calibration of $\vecDB$. Subsequently, we moved the magnet in the lateral $(x,y)$-plane of the laboratory frame. The direction and step size was determined from a field map of the permanent magnet and the residual transverse components of the field $\vecB$. We terminated this iterative process when the residual transverse field components were smaller than $50\,\mu\mathrm{T}$.

\textit{Determination of hyperfine couplings ($\apar,\aperp$) from ($f_0,f_{-1},f_R$):}
The hyperfine couplings $\apar$ and $\aperp$ in the limit $2\pi f_0 \gg \apar,\aperp$ are given by:
\begin{eqnarray}
\apar/(2\pi)  & = & f_{-1}-f_0 \\
\aperp/(2\pi) & = & \pi f_R
\end{eqnarray}
In our experiments the hyperfine couplings and the nuclear Larmor frequency $f_0$ were of similar magnitude, and we used the following transformations \cite{boss16} to obtain the hyperfine couplings. 
\begin{align}
\apar = 2\pi f_{-1} \left(\frac{\cos\left(2\pi f_{-1} \frac{\tau}{2}\right) \cos\left(2\pi f_0 \frac{\tau}{2}\right)-\cos(\pi-2\pi f_R \tau)}{\sin\left(2\pi f_{-1} \frac{\tau}{2}\right) \sin\left(2\pi f_0 \frac{\tau}{2}\right)}\right)-2\pi f_0
\end{align}
\begin{align}
\aperp = \sqrt{\left(2\pi f_{-1} \right)^2-\left(2\pi f_0 +\apar\right)^2} 
\end{align}

\textit{Monte Carlo error propagation:}
Confidence intervals for $\phi$ and $\aiso$ were obtained using a Monte Carlo method, as described in \cite{alper90}, which takes calibration uncertainties in the external fields $\vecB,\vecDB$ and in the observed precession frequencies $f_{m_s}$ into account. All parameters subject to uncertainty were assumed to follow a normal distribution. Precession frequencies were determined using a non-linear, least-squares fitting algorithm and their measurement uncertainties were obtained from the fit error \cite{boss16}. The uncertainty in the magnetic field components was estimated from the residuals between calculated and measured ODMR lines in the calibration method for $\vecB,\vecDB$, described before.

\textit{Nuclear g-factor enhancement:}
The nuclear g-factor enhancement factor $\alpha(\ms)$ given in Eq. (\ref{eq:gfactor}) of the main text is based on the approximation of small external bias fields $D\gg \ye B_0$. More generally the $\ms$-dependent enhancement factors are given by \cite{maze10}:
\begin{align}
\alpha(\ms) &= \frac{(3|\ms|-2)D+\ms \ye B_0}{D^2-\ye^2 B^2_0} \frac{\gamma_e}{\gamma_n}  \begin{pmatrix}
    A_{xx} & A_{xy} & A_{xz} \\
    A_{yx} & A_{yy} & A_{yz} \\
    0	      & 0          & 0 
\end{pmatrix},
\end{align}
which is also valid in the limit of large magnetic fields $\ye B_0 \gg D$ and provides, in principle, more accurate theory values for small $B_0$. We have analyzed our experimental data using this expression and found deviations to Eq. (\ref{eq:gfactor}) that are smaller than the frequency resolution in our experiments. 



%

\end{document}